\documentclass[aps,superscriptaddress,10pt,letterpaper,twocolumn]{revtex4}

\usepackage{graphicx}
\usepackage{amsmath}
\usepackage{amssymb}

\begin{document}

\title{Quantum optical coherence can survive photon losses:\\
a continuous-variable quantum erasure correcting code}

\author{Mikael Lassen$^{1}$}
\noaffiliation{}

\author{Metin Sabuncu$^{1,2}$}
\noaffiliation{}

\author{Alexander Huck$^1$}
\noaffiliation{}

\author{Julien Niset$^{3,4}$}
\noaffiliation{}

\author{Gerd Leuchs$^{2,5}$}
\noaffiliation{}

\author{Nicolas J. Cerf$^3$}
\noaffiliation{}

\author{Ulrik L. Andersen$^{1}$}
\noaffiliation{}

\affiliation{Department of Physics, Technical University of Denmark, Fysikvej, 2800 Kongens Lyngby, Denmark\\
$^{2}$ Max Planck Institute for the Science of Light, G\"{u}nther Scharowskystrasse 1, 91058 Erlangen, Germany\\
$^{3}$ Quantum Information and Communication, Ecole Polytechnique, CP 165, Universit´e Libre de Bruxelles, 1050 Brussels, Belgium\\
$^{4}$ Department of Physics, Hunter College of CUNY, 695 Park Avenue, New York, NY 10065\\
$^{5}$University Erlangen-N\"{u}rnberg, Staudtstrasse 7/B2, 91058 Erlangen, Germany.}

\begin{abstract}
\textbf{A fundamental requirement for enabling fault-tolerant quantum information processing is an efficient quantum error-correcting code (QECC)
that robustly protects the involved fragile quantum states from their environment~\cite{nielsen.book,shor1995.pra,steane1996.prl,calderbank1996.pra,laflamme1996.prl,bennett996.pra,grassl1997.pra,braunstein1998.prl,slotine1998.prl,braunstein1998.nat}.
Just as classical error-correcting codes are indispensible in today's information technologies, it is believed that QECC will play
a similarly crucial role in tomorrow's quantum information systems.
Here, we report on the first experimental demonstration of a quantum erasure-correcting code that overcomes the devastating effect of photon losses. Whereas {\it errors} translate, in an information theoretic language, the noise affecting a transmission line, {\it erasures}
correspond to the in-line probabilistic loss of photons. Our quantum code protects a four-mode entangled mesoscopic state of light
against erasures, and its associated encoding and decoding operations only require linear optics and Gaussian resources.
Since in-line attenuation is generally the strongest limitation to quantum communication, much more than noise,
such an erasure-correcting code provides a new tool for establishing quantum optical coherence over longer distances.
We investigate two approaches for circumventing in-line losses using this code, and demonstrate that
both approaches exhibit transmission fidelities beyond what is possible by classical means.}
\end{abstract}

\date{\today}

\maketitle

Quantum information protocols are inevitably affected by noise, which in turn produces errors in the extremely sensitive processed quantum information~\cite{nielsen.book}. Thus, in order to gain the full advantage of quantum information processing, including long-distance quantum communication and fault-tolerant quantum computing~\cite{duan2005.nat,yuan2008.nat,knill2005.nat}, these errors must be efficiently corrected. This can be done by encoding the information in special quantum error-correcting codes, which introduce redundancy and thereby protect the fragile quantum information from environment-induced decoherence. Using such codes, the transmission errors can be diagnosed through so-called syndrome measurements, the results of which are used to correct the corrupted quantum information.

Quantum error correcting codes (QECC) were first discovered for discrete variable qubit systems~\cite{shor1995.pra,steane1996.prl,calderbank1996.pra,laflamme1996.prl,bennett996.pra,grassl1997.pra} and later extended to systems where information is encoded into observables with a continuous spectrum~\cite{slotine1998.prl,braunstein1998.prl,braunstein1998.nat,wilde2007.pra,niset2008}.
Only a few experimental implementations demonstrating quantum error correction have been carried out to date, e.g., in nuclear magnetic resonance systems~\cite{cory1998.prl}, in an ion-trap system~\cite{Chiaverini2004.nat}, and in a pure optical system~\cite{Pittmann2005.pra,aoki2008.xxx}. All these works have reported on the correction of errors, which are the manifestation of line noise. However, it is very often the loss of photons in a transmission line (corresponding to {\it erasures} in an information theoretic language) that is the main obstacle to the survival of quantum coherence.

Erasure-correcting codes have long been known in classical coding theory, and their quantum counterparts have also been theoretically developed.
The experimental progress has been hampered by the experimental complexity involved in the original Shor-based continuous variable (CV) QECC that relies on a nine-mode entangled state~\cite{braunstein1998.nat,aoki2008.xxx}. In our manuscript, we report on the first experimental realization of a quantum {\it erasure}-correcting code which simultaneously protects two independent continuous-variable (CV) quantum systems against photon losses in the transmission channel.

We consider the transmission of a quantum state of light through a channel which either transmits the information perfectly or completely erases it with an error probability, $P_E$. The density matrix of the transmitted state is given by
\begin{equation}
\rho=(1-P_E)|\alpha\rangle\langle\alpha|+P_E|0\rangle\langle 0|,
\label{state}
\end{equation}
where the state $|\alpha\rangle$ comprises the transmitted quantum information and $|0\rangle$ is the vacuum state arising due to channel erasure. Such an error model, which can be viewed as random fading, is likely to occur as a result of e.g. time jitter noise or beam pointing noise in an atmospheric transmission channel~\cite{book,wittmann2008.pra,elser2009.njp}. The CV code for protecting quantum information from erasures is a four-mode entangled mesoscopic state in which two (information-carrying) quantum states are encoded with the help of a two-mode entangled vacuum state~\cite{niset2008}. As a result of the redundancy, the damage that the erasure has made onto the quantum states can be reversed by a simple decoding procedure followed by measurements that determine the damage and a feedforward step that corrects the output states. It is remarkable that the quantum coding, decoding and correction is based on simple linear optical components and Gaussian resources. This is not a violation of the recent No-Go theorem for QECC of Gaussian states with Gaussian transformation~\cite{niset2009} since the stochastic erasure noise occurring in the transmission channel produces a non-Gaussian state.

We investigate two different detection strategies for correcting the corrupted quantum states. The first strategy actively corrects the errors of the transmitted state according to the outcome of the measurement, whereas the second strategy filters out the corrupted state when an error is detected. A fundamental difference between the two approaches is that the former is deterministic whereas the latter is probabilistic. As for most QECC schemes, the deterministic approach comes with a price: For the scheme to work one may only allow for the occurrence of a given number of errors (a single error in this case) and one needs to know in which mode the error occurred. As was shown in Ref.~\cite{niset2008} these requirements may, however, be relaxed by using a probabilistic approach. We note that a subpart of the deterministic circuit has been implemented in the context of CV quantum secret sharing~\cite{Lance}.

\begin{figure}[t]
\begin{center}
\includegraphics[width=0.45\textwidth]{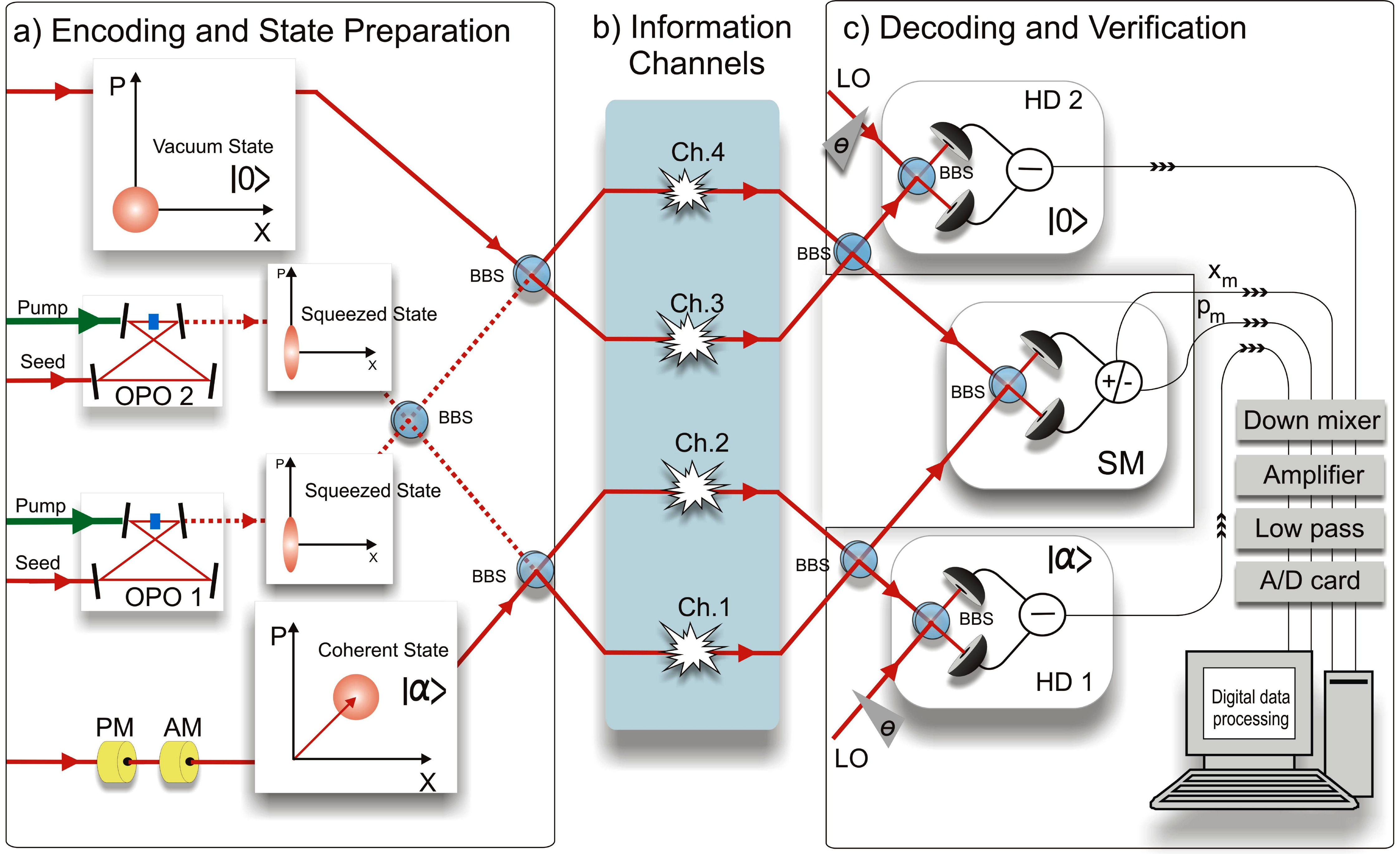}
\caption{\textbf{Schematics of the experimental QECC setup.} a) The four-mode code is prepared through linear interference at three balanced beam splitters (BBS) between the two input states, $|\alpha\rangle$ and $|0\rangle$, and two ancillary squeezed vacuum states. The latter states are produced in two optical parametric oscillators, OPO 1 and OPO 2, and the coherent state is prepared via a coherent modulation at 5.5 MHz produced by an amplitude (AM) and a phase modulator (PM).
b) The encoded state is injected into four free-space channels that can be independently blocked, thereby mimicking erasures. c) The corrupted state is decoded, the error is detected by the measurement (M) and the state is deterministically corrected or probabilistically selected. The measurement is an entangled measurement where the phase and amplitude quadratures of the two emerging states are jointly measured (see e.g Ref.~\cite{niset2007.prl}).
The error correcting displacement or post-selection operation is carried out electronically after the measurement of the transmitted quantum states. These states are measured with two independent homodyne detectors which allow for full quantum state characterization, by scanning the phases ($\theta$) of the local oscillators (LO) with respect to the phases of the signals. All erasure events are obtained by blocking the beam paths.}
\label{setup-Scheme}
\end{center}
\end{figure}

The schematics of our setup is depicted in Fig.~\ref{setup-Scheme}. It consists of an encoding station where the four-mode code is prepared, an erasure channel where information is randomly erased and a decoding station, where measurement outcomes either correct or filter the corrupted state. The key element in the preparation stage is a two-mode Gaussian entangled source which exhibits quantum correlations between pairs of conjugate quadrature amplitudes. The two-mode squeezing (or CV entanglement) is produced through interference of two single-mode squeezed states generated in optical parametric oscillators (OPOs) operating below threshold. The main source for the OPOs, is a Diabolo laser from Innolight delivering 400 mW of infra-red light (1064 nm) and 600 mW of green light (532 nm). These light beams are both sent through empty, high-finesse triangle-shaped cavities in order to clean their spatial modes as well as to filter out classical amplitude and phase noise. The resulting beams serve as pump beams, as seed beams and as locking beams for the OPOs. The OPOs are bow-tie shaped cavities each with a type I periodically poled Potassium Titanyl Phosphate crystals (1x2x10 mm$^3$) for nonlinear down conversion. The OPO cavities consist each of two curved mirrors of 25 mm radius of curvature and two plane mirrors. Three of the mirrors in each cavity are highly reflective at 1064~nm ($R>99.95$\%) while the output couplers have a transmission of 8\%. The transmittance of the mirrors at the pump wavelength (532~nm) is more than 95$\%$. Along with the pump we inject a seed beam at 1064 nm, the brightness of which is facilitating the construction of the various phase locks in the setup. To lock the phase of the cavities we use auxiliary counter propagating beams at 1064nm. Amplitude quadrature squeezed beams are then produced through de-amplification of the seed beams. Using a spectrum analyzer we measure squeezing of $-3.4\pm0.2$~dB and $-2.7\pm0.2$~dB below the shot noise level (at 5.5~MHz and a resolution bandwidth of 300 kHz and video bandwidth of 300 Hz) for the two OPOs. The two amplitude squeezed beams with a relative phase of $\pi$/2 are then brought to interference at a 50/50 beam splitter (visibility greater than 98\%) to produce the two-mode squeezing which is then launched into the quantum error correction coding setup. The average measured two-mode squeezing is approximately $-2.0\pm0.5$~dB below the shot noise level.

This state interferes with two signal states to form the final four-mode code comprising four optical beams. A vacuum state is chosen as one of the input signals, while the other input is prepared
in a coherent state. This choice simplifies the experimental realization, but is not an intrinsic limitation of the scheme. The four resulting beams are then dispersed into four free space transmission channels which can be independently blocked to simulate any combination of erasures. At the decoding station the interferences are reversed in two balanced beam splitters, and two of the resulting outputs are jointly measured in an entangled measurement strategy~\cite{niset2007.prl}, where the modes interfere on a balanced beam splitter and conjugate quadratures are measured at the two outputs (see Fig.~\ref{setup-Scheme}). The resulting outcomes are now used either to deterministically correct the errors through conditional linear displacements or to probabilistically filter out the loss-infected states.

\begin{figure}[t]
\begin{center}
\includegraphics[width=0.45\textwidth]{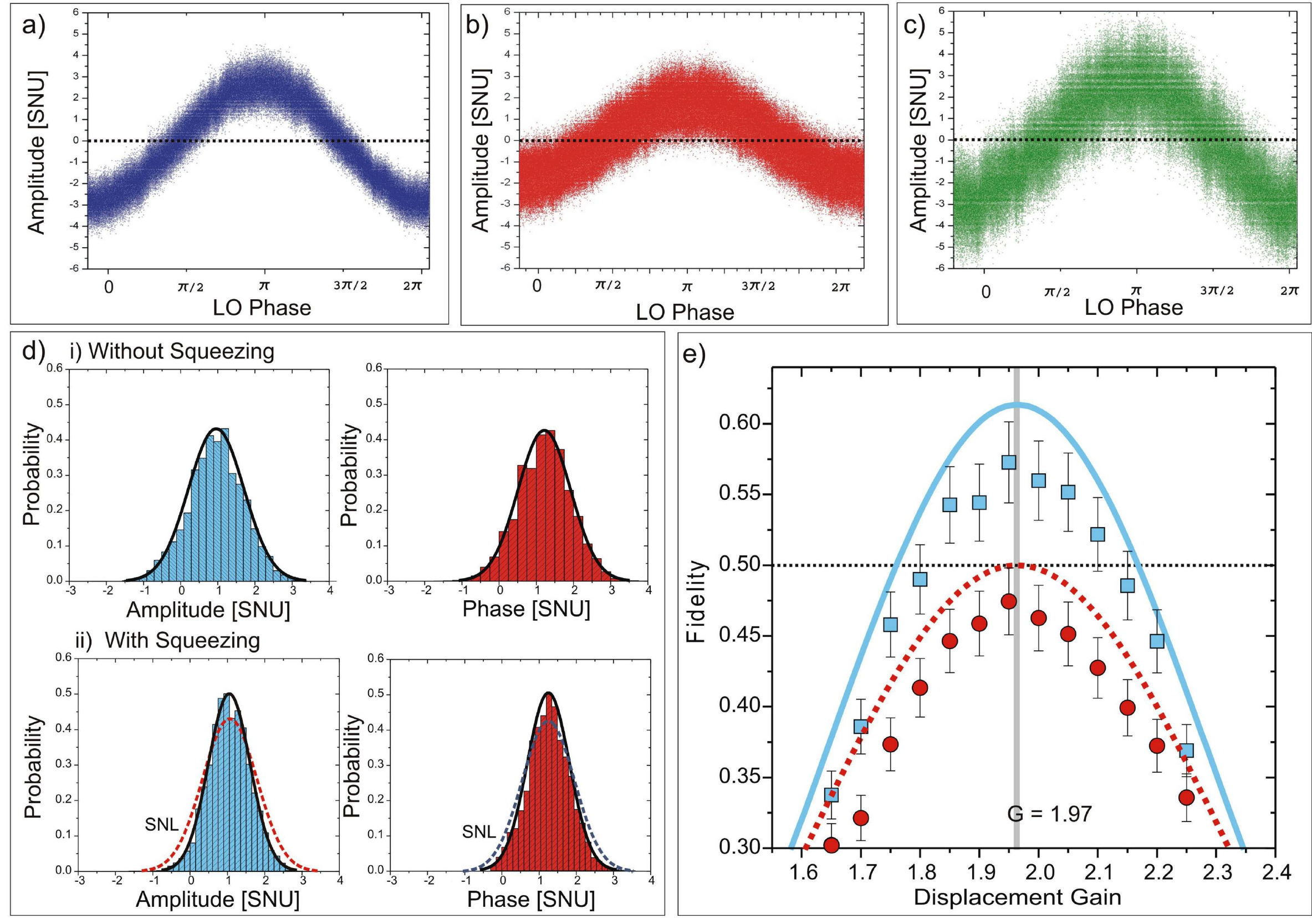}
\caption{\textbf{Results of the deterministic QECC protocol.} a) Phase scan of the input coherent state with the excitation $|\alpha\rangle\approx|3+3i\rangle$. b) Phase scan of the output state measured at HD1 before correction and c) Phase scan of the corrected output state. d) Histograms of the marginal distributions of the amplitude and phase quadratures of the joint syndrome measurement (in shot noise units (SNU)). The red curves correspond to the marginal distributions for a shot noise limited (SNL) state whereas the black curves are the best Gaussian fits to the histograms. e) Fidelity is plotted as a function of the displacement gain with the use of entanglement (blue squares) and without the use of entanglement (red circles). The dashed and solid lines are the theoretically predicted fidelities for 0 dB and 2 dB of two-mode squeezing, respectively.}
\label{Fig2}
\end{center}
\end{figure}

All measurements are performed at a sideband frequency of 5.5~MHz. The electronic output of all detectors are mixed down with an electronic local oscillator at 5.5~MHz, low-pass filtered (300 KHz), amplified (FEMTO DHPVA-100) and finally digitized by a 14 bit A/D converter at 10 Msamples/s. The signal is sampled around the 5.5MHz sideband to avoid low frequency classical noise. The time trace of the coherent state is shown in Fig.~\ref{Fig2} before (a) and after (b) the total erasure of Ch.2. These are typical traces of a homodyne measurement of a coherent state, where the local oscillator is scanned from 0 to  2$\pi$. The traces consist of approximately 220000 data points. From the down-mixed time traces we reconstruct the density matrices and the Wigner functions using a maximum likelihood algorithm.

First we describe the deterministic correction strategy. Fig.~\ref{Fig2}a shows a scan of the quantum mechanical oscillator comprising the coherent state quantum information of the input state. The information encoded in a coherent state can be concisely described by the conjugate quadrature operators; the amplitude $\hat x$ and the phase $\hat p$ such that $|\alpha\rangle=|\langle \hat x\rangle+i\langle \hat p\rangle\rangle$. For the specific measurement run shown in Fig.~\ref{Fig2}a, $|\alpha\rangle\approx|3+3i\rangle$. Fig.~\ref{Fig2}b illustrates the measurements at the homodyne detector HD1 after the four-mode state has been transmitted through the channel with erasure on channel 2. The state is clearly seen to be corrupted as the first and second moments of the quantum oscillator are significantly changed.
However, by using the measurement outcomes ($x_m,p_m$) of the homodyne detectors (see Fig.~\ref{Fig2}c) to linearly displace the amplitude and phase quadratures of the transmitted state ($x_o,p_o$), the displacement is given by  
\begin{eqnarray}
x_o\rightarrow x_o+\sqrt{G}x_m\\ 
p_o\rightarrow p_o+\sqrt{G}p_m,\nonumber
\end{eqnarray}
where $G$ is the displacement gain. The original quantum state is partially recovered as shown qualitatively in Fig.~\ref{Fig2}d for $G=1.97$. The accuracy in the estimation of the error (and thus the precision of the displacement) depends crucially on the degree of squeezing: By employing infinite squeezing, the transmitted states can, in principle, be perfectly corrected \cite{niset2008}.
With finite squeezing the protocol can be quantified by the fidelity between the input state and the corrected output state. Based on the measurements presented above, the fidelities are computed for various gains and the results are depicted by the blue squares in Fig.~\ref{Fig2}e. A maximum fidelity of about 0.57$\pm0.2$ is obtained which clearly surpasses the classical benchmark of 0.50. Similar fidelities are achieved for the erasure of channel 1, whereas fidelities close to unity are obtained when channel 3 or 4 are blocked. Measurements for which the two-mode squeezed state was replaced by vacua are also carried out for different displacement gains. The resulting fidelities are depicted in Fig.~\ref{Fig2}e by the red circles, and they nicely illustrate the need for entanglement.

\begin{figure}[t]
\begin{center}
\includegraphics[width=0.45\textwidth]{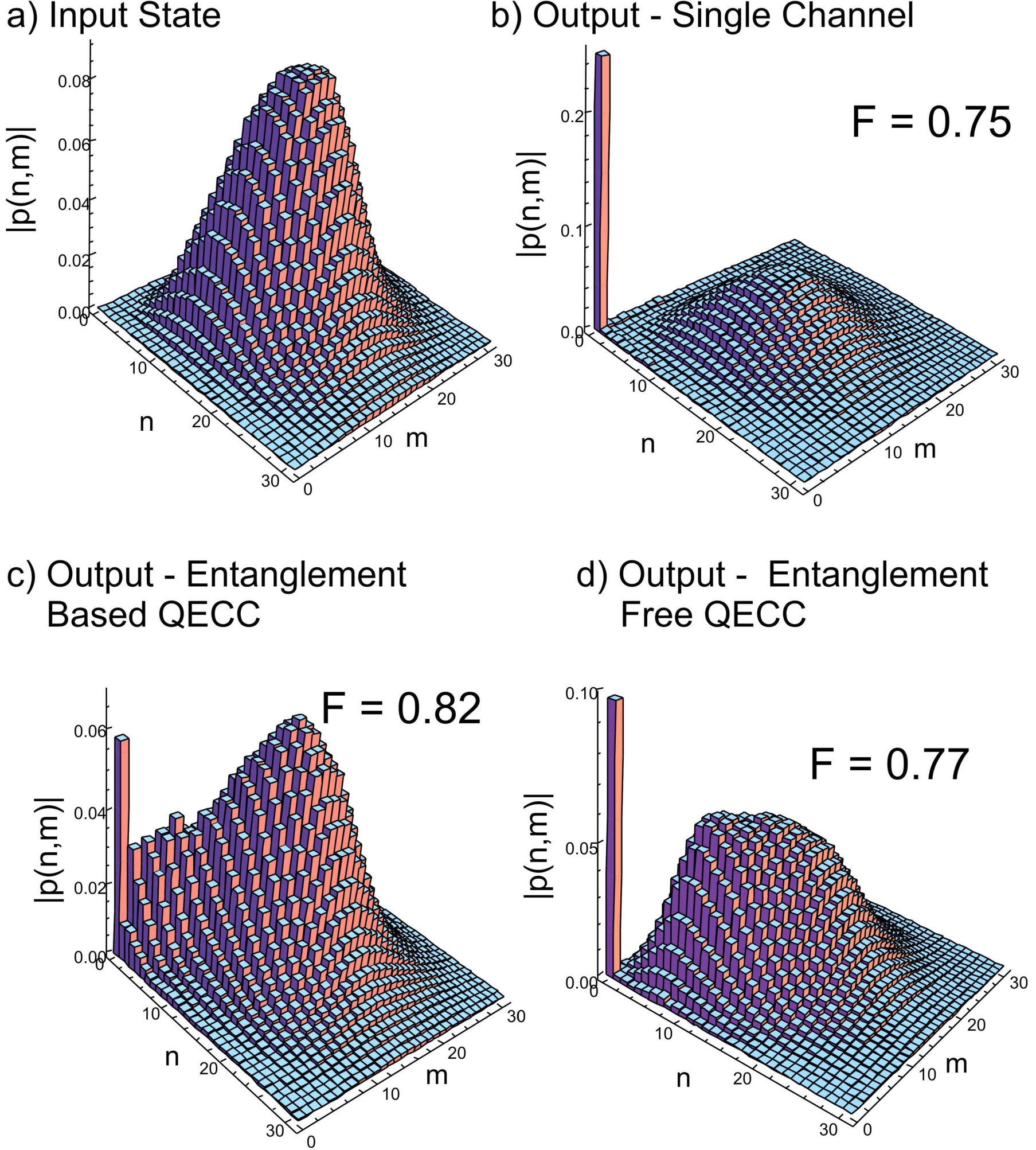}
\caption{\textbf{Results of the probabilistic QECC protocol.} a) Density matrix in the 30$\times$30 Fock state basis of the input coherent state with the excitation $|\alpha\rangle\approx|3+3i\rangle$. b) Density matrix of the output state using a single erasure channel as described in eq. (\ref{state}). The fidelity to the input state is $F=75\%$. c) Density matrix of the corrected output state employing the entangled four-mode code. The fidelity to the input state is $F=82\%$. d) Density matrix for the corrected output state when the entangled states are blocked, and thus replaced by vacua. The fidelity is computed to $F=77\%$. The erasure probability for all realizations is $P_E=0.25$.}
\label{Fig3}
\end{center}
\end{figure}

We now proceed by discussing the results of probabilistic recovery of quantum information. In contrast to the deterministic approach where stringent conditions were put on the channels, the probabilistic approach is much less stringent: One may allow for multiple erasures and, in addition, there are no requirements on the knowledge of the occurrence and location of the erasures. In this case, the density matrix of the transmitted state is given by
\begin{equation}
\rho_{trans}=\sum_{i=1}^{16} P_i \rho_i,
\label{rho}
\end{equation}
where $\rho_{i}$ is the output density matrix corresponding to one of the sixteen different erasure patterns that may occur during transmission and the $P_i$'s are the associated probabilities running from $P_E^4$ to $(1-P_E)^4$ corresponding to complete erasure and complete transmission, respectively. Using the tomographic maximum likelihood recipe for reconstructing the density matrices via homodyne detection, we fully characterized the input and output states for various cases. Fig.~\ref{Fig3}a shows the density matrix of the coherent input state in the 30$\times$30 Fock state basis. This state is then mixed with the entangled state and sent through the four channels. Subsequently, we perform 16 different full measurement runs by interchangeably blocking the four channels corresponding to the 16 different transmission patterns. The measurement outcomes are then weighted by the probabilities $P_i$ in order to create the density matrix of the transmitted mixed state. The corrupted states are then probabilistically corrected by conditioning on the outcomes of the SMs: For the realization in Fig.~\ref{Fig3} we used the condition that if the measurement outcomes obeyed $|x_m|>0.8$ and $|p_m|>0.8$ (found from optimization~\cite{niset2008}), an error was detected and the resulting transmitted state was discarded. After this filtering operation, we reconstruct the density matrix based on the reduced data set and the result is shown in Fig.~\ref{Fig3}c for $P_E=0.25$. By replacing the entangled states with vacua, the resulting density matrix is largely changed as illustrated in Fig.~\ref{Fig3}d. To determine the fidelity, $F$, between the input state (with density matrix $\rho_{in}$) and the filtered output state (with density matrix $\rho_{out}$) we use the general expression
\begin{eqnarray}
\text{F}=[\text{Tr}(\sqrt{\sqrt{\rho_{out}}\rho_{in}\sqrt{\rho_{out}}})]^2.
\end{eqnarray}
For the example in Fig.~\ref{Fig3}c we compute a fidelity of $F=0.82\pm 0.02$ which clearly surpasses the transmission fidelity of $F\simeq 0.75$ obtained by transmission in a single channel with similar erasure probability.
It is interesting to note that even without entanglement, the protocol also outperforms the single channel approach. For the corresponding entanglement-free setup, were the two-mode squeezed state is replaced by vacua, the corrected density matrix is shown in Fig.~\ref{Fig3}c and we compute a fidelity of $F=0.77\pm 0.02$.

\begin{figure}[t]
\begin{center}
\includegraphics[width=0.45\textwidth]{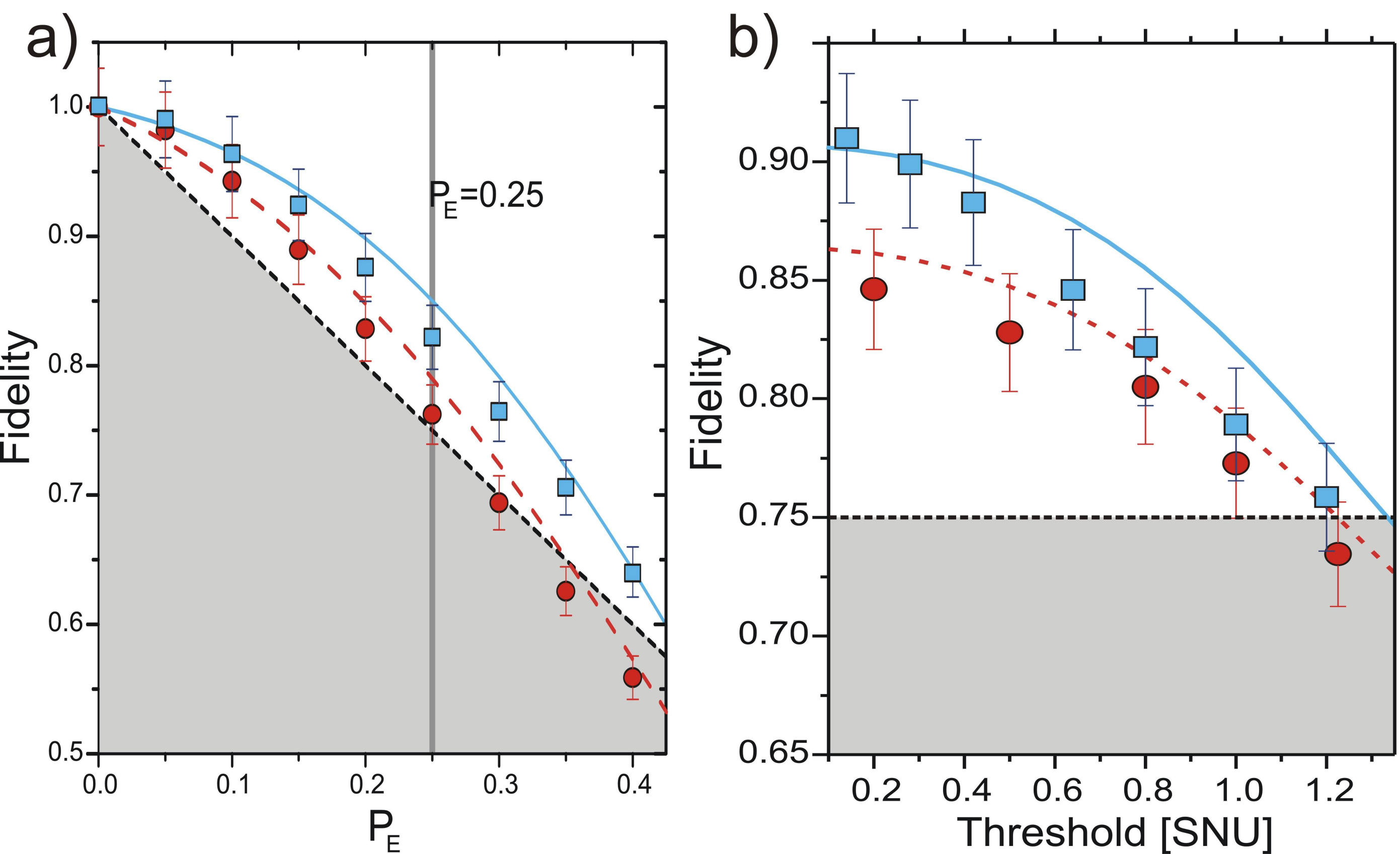}
\caption{\textbf {Performance of the probabilistic QECC protocol.}
Transmission fidelity is plotted as a function of the erasure probability (in a) and the postselection threshold value (in b). In both plots the dashed line represents the single channel (code-free) fidelity. The circles correspond to the entanglement-free error correction code, whereas the squares correspond to the entanglement based code. The dashed and solid lines are the theoretically predicted fidelities for the two cases. The directly measured squeezing values at the entrance to the protocol were around 3dB (see methods). The black dotted lines are the fidelity for the transmission in a single channel.
}\label{Fig4}
\end{center}
\end{figure}

The fidelity as a function of the error probability $P_E$ and the threshold value $|x_{th}|=|p_{th}|$ are illustrated in Fig.~\ref{Fig4}a and \ref{Fig4}b, respectively. We see that the entanglement-free QECC performs better than the single channel approach (i.e. without error correction) for error probabilities up to about 28\%. It is also evident that the use of the entanglement-based code further increases the fidelity for all evaluated error probabilities. As expected, we see in Fig.~\ref{Fig4}b that the fidelity increases as the threshold value is lowered. This is associated, however, with a lower success probability.

We have successfully demonstrated a continuous-variable quantum erasure-correcting code that protects quantum optical coherence from erasures (or probabilistic photon losses). The deterministic version of our scheme has the additional advantage that it enables a direct
processing of the error-corrected output state in downstream applications,
circumventing the need for a quantum repeater configuration that requires
quantum entanglement distillation and quantum memory. Although the protocol has only been experimentally accessed for quantum information encoded in coherent states of light, it enables the faithful transmission of other quantum states such as squeezed states, qubit states or even bipartite entangled states. Furthermore, the error model can be extended to include also partial erasure and random phase noise, which often occurs in free space transmission in a turbulent atmosphere. As an outlook, it is intriguing to address the universality of our protocol with respect to arbitrary input states and arbitrary non-Gaussian error models.

This work was supported by the Future and Emerging Technologies programme of the European Commission under the FP7 FET-Open grant no. 212008 (COMPAS),
by the Danish Agency for Science Technology and Innovation under grant no. 274-07-0509, by the Deutsche Forschungsgesellschaft, and by the Interuniversity Attraction Poles program of the Belgian Science Policy Office under grant IAP P6-10 (photonics@be).

\end{document}